\documentclass[a4paper,11pt]{article}
\pdfoutput=1 

\usepackage{jheppub} 
\usepackage{hyperref}

\usepackage[T1]{fontenc} 
\usepackage{amssymb,amsmath,graphicx,amscd,xcolor,amsthm}
\usepackage{braket}
\usepackage{verbatim} 

\def\x{\mathbf{x}}

\title{\boldmath Remarks on the influence of quantum vacuum fluctuations over a charged test particle near a conducting wall}
\author[a]{V. A. De Lorenci,}
\author[b]{C. C. H. Ribeiro,}

\affiliation[a]{Instituto de F\'{\i}sica e Qu\'{\i}mica,  Universidade Federal de Itajub\'a,\\
Itajub\'a, Minas Gerais 37500-903, Brazil}
\affiliation[b]{Instituto de F\'{\i}sica de S\~ao Carlos, Universidade de S\~ao Paulo,\\ S\~ao Carlos, S\~ao Paulo 15980-900, Brazil}

\emailAdd{delorenci@unifei.edu.br}
\emailAdd{caiocesarribeiro@ifsc.usp.br}

\date\today

\abstract{
Quantum vacuum fluctuations of the electromagnetic field in empty space seem not to produce observable effects over the motion of a charged test particle. However, when a change in the background vacuum state is implemented, as for instance when a conducting boundary is introduced, dispersions of the particle velocity may occur. As a consequence, besides the existence of classical effects due to the interaction between  particle and  boundary, there will be a quantum contribution to the motion of the particle whose magnitude depends on how fast the transition between the different vacuum states occurs.
Here this issue is revisited and a smooth transition with a controllable switching time between the vacuum states of the system is implemented.
Dispersions of the particle velocity in both, zero and finite temperature regimes are examined.
More than just generalizing previous results for specific configurations, new effects are unveiled. Particularly, it is shown that the well known vacuum dominance reported to occur arbitrarily near the wall is a consequence of assumed idealizations. The use of a controllable switching enable us to conclude that thermal effects can be as important as or even stronger than vacuum effects arbitrarily near the wall.
Additionally, the residual effect predicted to occur in the late time regime was here shown to be linked to the duration of the transition. In this sense, such effect is understood to be a sort of particle energy exchanging due to the vacuum state transition.
Furthermore, in certain arrangements a sort of cooling effect over the motion of the particle can occur, i.e., the kinetic energy of the particle is lessen by a certain amount due to subvacuum quantum fluctuations.
}

\keywords{Boundary quantum field theory, Quantum vacuum fluctuations, Stochastic processes}

\begin{document} 
\maketitle
\flushbottom

\section{Introduction}
Vacuum fluctuations of the electromagnetic field in the presence of a perfectly conducting flat wall affect the motion of a nearby charged test particle.
This system was originally investigated \cite{ford2004,ford2005} using a model in which the interaction is suddenly turned on at a given initial time, and after an interval of time $\tau$ the dispersion of the particle velocity is thus calculated.
Because of the idealizations assumed in the description of the system, the dispersions are plagued with some divergences. 
%
%
%
Several aspects about this system were discussed in the literature, as for instance, the generalization to the case with two reflecting walls \cite{hongwei2004}, the behaviour of the dispersions when finite temperature effects are included \cite{hongwei2006}, and the implementation of switching mechanisms \cite{seriu2008,seriu2009,delorenci2016}, among others. 
Particularly, it was shown \cite{delorenci2016} that the assumption of a smooth switching connecting the two distinct states of the system -- the particle in empty space and in the presence of a conducting wall -- is enough to regularize all divergences found in the original model.
An important feature exhibited by this system is the occurrence of a residual effect, i.e., in the late time regime the particle at a distance $z$ from the wall presents a non null constant dispersion of its velocity. The existence of such a residual effect was suggested to be linked with an energy conservation law \cite{ford2004}. This issue was further discussed in a toy model based on a real massless scalar field \cite{camargo2018}, where the contribution to the kinetic energy of the test particle was calculated under certain assumptions.

Here, in order to bring more reality to the description of this system, a smooth transition with a controllable switching time between the two vacuum states is implemented.  Previous analyses \cite{ford2004,hongwei2004,delorenci2016} were generalized, and new effects unveiled.
Because of the existence of a natural scale of distance in this model, several aspects about the behaviour of the dispersions of the particle velocity in the regime of small distances could be elucidated.
Late time regime is also discussed and it is shown that the magnitude of the residual effect over the motion of the particle depends on the duration of the transition.
Additionally to the classical effects due to the interaction between the charged particle and the wall, a change in its kinetic energy sourced by the switched vacuum state takes place. Such quantum contribution to the energy of the particle can be positive or negative, depending on how fast the vacuum transition is implemented. For fast transitions the particle gains a certain amount of energy, while for slow transitions the particle somehow loses part of its kinetic energy as a consequence of subvacuum quantum fluctuations.
Finite temperature effects were also examined. Similarly to the case of zero temperature, the divergences appearing in the dispersions when a sudden switching is implemented \cite{hongwei2006} are naturally regularized when a smooth transition is implemented.
One interesting aspect unveiled by the implementation of the smooth switching is that thermal effects over the motion of the particle can be comparable to, or even greater than, vacuum effects arbitrarily near the boundary.

Regarding the above mentioned subvacuum fluctuations, they consist of phenomena related to the occurrence of negative quantum expectation values of quantities that are positive defined in the realm of classical physics. For instance, the renormalized vacuum expectation value of the squared electric field in a squeezed vacuum state can be negative for certain configurations, leading to several consequences on the propagation of light in a nonlinear optical material \cite{bessa2014,delorenci2019}. As another example, negative energy density in Casimir-like systems and its gravitational consequences has been investigated in the literature \cite{sopova2002,sopova2005}. It is interesting to notice that the duration in which subvacuum effects can exist is constrained by certain quantum inequalities \cite{ford1978,ford1991,ford1997,flanagan1997,fewster1998}.

Next section addresses the expression describing the dispersion of the velocity of an electric charged particle in the presence of a reflecting wall. It also includes the choice of the switching function that makes a smooth connection between the two distinct vacuum states. 
Assuming a scenario of zero temperature, dispersions of the velocity parallel and perpendicular to the wall are calculated in Sec.~\ref{dispersions}, and their behaviours discussed. Particularly, some aspects exhibited by these dispersions in the late time regime and near the wall are addressed. 
Furthermore, the change in the energy of the particle due to the transition between the vacuum states is also discussed. 
It is shown that the kinetic energy of the particle can be reduced due to subvacuum fluctuations.
A scenario of finite temperature is adopted in Sec.~\ref{finite}, where the dispersions are calculated and their behaviours discussed. Previous results are recovered, typical divergences are regularized, and new effects near the wall are unveiled.
Vacuum to thermal dominance near the wall is also investigated.
Final remarks are presented and Sec.~\ref{final}, including a brief discussion connecting the results obtained in this work with previous results reported in the literature. 
%
Appendix \ref{appendix} presents additional steps that are useful in the calculation of the dispersions obtained in Sec.~\ref{dispersions}. Finally, thermal contributions to the dispersions at the late time regime are presented in appendix \ref{appendixii}.

Our analysis is restricted to the case where spin degree of freedom of the charged particle can be neglected. Otherwise explicitly stated, units are such that $\hbar = c =1$. Additionally the vacuum dielectric permittivity $\epsilon_0$ and the Boltzmann constant $ \kappa_{{}_B} $ are set to unit, which makes $1{\rm V} \approx 1.67\times 10^7 {\rm m}^{-1}$  and $1{\rm K} \approx 4.37\times 10^2 {\rm m}^{-1}$ in our units.

\section{Preliminaries}
\label{preliminaries}
We start by summarizing the main aspects about the system we would like to study. Suppose that a non relativistic test particle of mass $m$ and electric charge $q$, initially in empty space, experiences a smooth transition to a space containing an infinity perfectly conducting flat wall, which is placed at $z=0$. 
Thus, the equation of motion of the particle interacting with a background quantum electric field prepared in its vacuum state in the presence of the wall, leads to the following expression governing the dispersion (mean squared deviation) of the $j$-th component of the particle velocity \cite{delorenci2016},
\begin{equation}
\langle(\Delta v_j)^2\rangle = \frac{q^2}{m^2}\int_{-\infty}^\infty\int_{-\infty}^\infty \langle E_j(\x,t)E_j(\x,t')\rangle_{{}_{\tt vacuum}}F_{\tau_s,\tau}(t)F_{\tau_s,\tau}(t')dtdt',
\label{i1}
\end{equation}
where $F_{\tau_s,\tau}(t)$ is a switching function that allows a smooth transition between the two scenarios above described, and $\langle E_j(\x,t)E_j(\x,t')\rangle_{{}_{\tt vacuum}}$ denotes the $j$-th component of the renormalized electric field correlation function (positive Wightman two-point function). Here, renormalization means that the unbounded Minkowski space (hereafter called empty space) contribution was subtracted. The parameter $\tau$ denotes the interval of time for which the particle is effectively interacting with the quantum field in the presence of the conducting wall, while $\tau_s$ (switching time) is defined as the interval of time in which the system performs the transition between the two scenarios -- from empty-space to the presence of a conducting wall, or vice versa. 
The condition of small particle displacement \cite{ford2004,delorenci2016} is here being used, i.e., the particle position $\x$ is assumed to be nearly a constant. 

The model exhibiting a sudden transition \cite{ford2004} can be exactly described by setting $F_{\tau_s,\tau}(t) \rightarrow \Theta(t)\Theta(\tau-t)$, where the unit step function $\Theta(t)$ is equal to 0 for $t < 0$, and 1 for $t \ge 1$. 
In such idealized case the leading contribution for the residual dispersion ($\tau/z \gg 1$) of the particle velocity perpendicular to the wall is given by
%
%
$\langle\left(\Delta v_{{}_\perp}\right)^2\rangle\approx q^2/(4 \pi ^2 m^2 z^2)$,
while in the parallel directions $\langle(\Delta v_{{}_\parallel})^2\rangle \approx -q^2/(3 \pi ^2 m^2 \tau ^2)$, which vanishes when $\tau\rightarrow\infty$. Throughout the text we denote $  \langle(\Delta v_{{}_\parallel})^2\rangle  \doteq \langle(\Delta v_x)^2\rangle=\langle(\Delta v_y)^2\rangle$ and  $ \langle(\Delta v_{{}_\perp})^2\rangle   \doteq \langle(\Delta v_z)^2\rangle$.

\begin{figure}[h!]
\center
\includegraphics[scale=0.38]{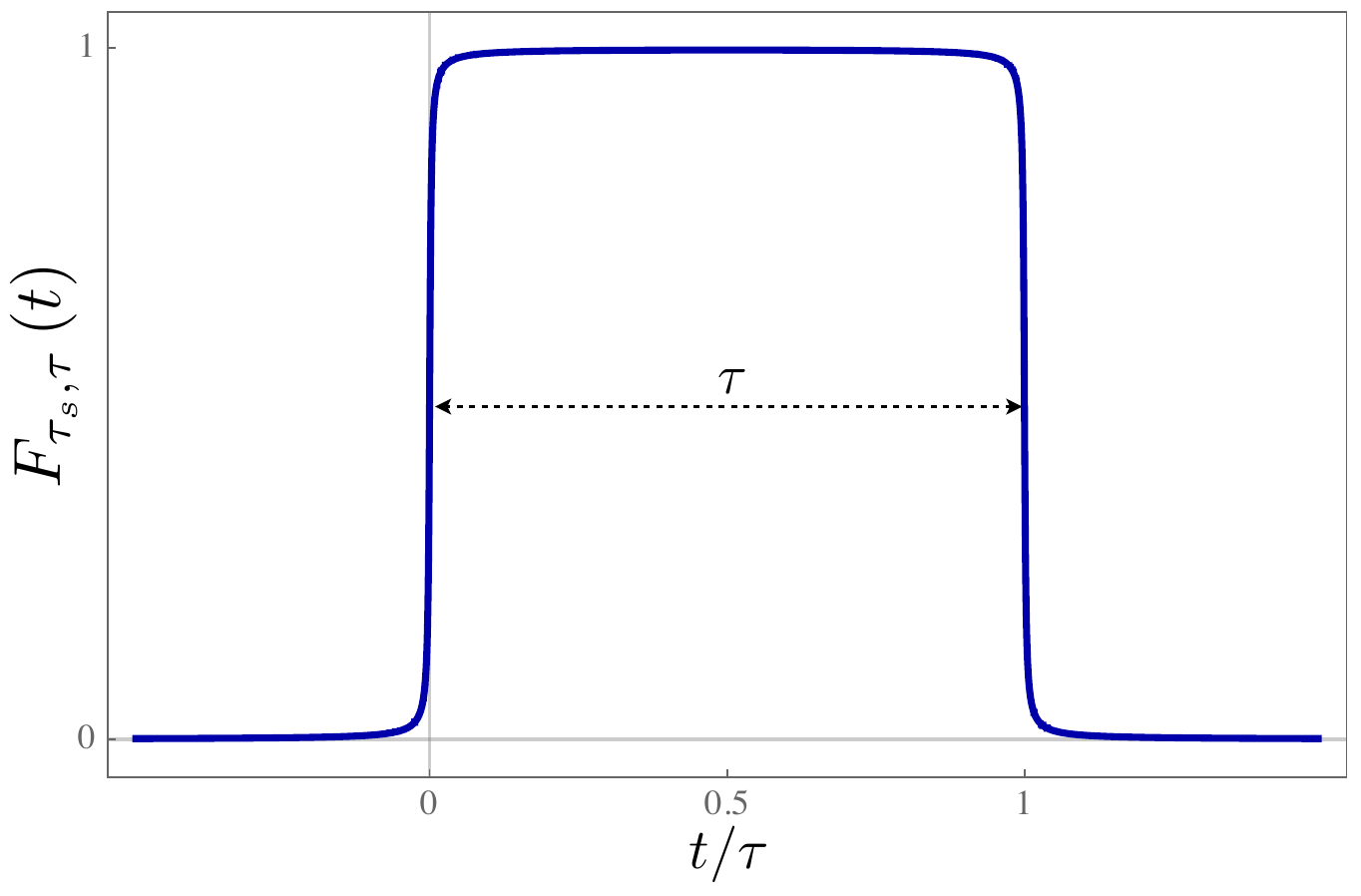}
\caption{Time behaviour of the switching function $F_{\tau_s,\tau}\left(t\right)$ defined by Eq.~(\ref{int27}), with a switching time such that $\tau_s/\tau = 0.002$.}
\label{fig0}
\end{figure}

Let us implement a smooth switching described by the function $F_{\tau_s,\tau}\left(t\right)$ defined as \cite{bessa2016},
\begin{equation}
F_{\tau_s,\tau}\left(t\right)=\frac{1}{\pi}\left[\arctan{\left(\frac{t}{\tau_s}\right)}+\arctan{\left(\frac{\tau-t}{\tau_s}\right)}\right],
\label{int27}
\end{equation}
where $\tau$ is approximately the width of the switching function, and $\tau_s$ measures the duration of the switching, which corresponds to the nearest interval between two successive points of maximum curvature in this function. 
The behaviour  of this function, as well as the geometric meaning of $\tau$ and $\tau_s$, is described in Fig.~\ref{fig0}, where we chose $\tau_s/\tau = 0.002$. 
The advantage of implementing such switching is that even in a late time regime the switching duration $\tau_s$ can be chosen to be finite, which is associated with a non-adiabatic process.  The idealized sudden switching is obtained in the limit $\tau_s\rightarrow 0$, for a fixed $\tau$. An adiabatic transition, i.e., a process that takes an infinite duration, is obtained by setting $\tau_s\sim\tau\to\infty$.

\section{Zero temperature regime}
\label{dispersions}
\subsection{Dispersions at zero temperature}
\label{zero-dispersions}
The renormalized vacuum correlation functions in Eq.~(\ref{i1}) are obtained by quantizing the electric field in the presence of the reflecting boundary and subtracting the Minkowski vacuum contribution \cite{brown1969}. The relevant correlations to our purposes are $\langle E_x(\x,t)E_x(\x,t')\rangle_{{}_{\tt vacuum}} = \langle E_y(\x,t)E_y(\x,t')\rangle_{{}_{\tt vacuum}} 
= -\pi^{-2}\left[(\Delta t)^2 +4z^2\right][(\Delta t)^2-4z^2]^{-3}$, and $\langle E_z(\x,t)E_z(\x,t')\rangle_{{}_{\tt vacuum}} =\pi^{-2}[(\Delta t)^2-4z^2]^{-2}$, where the limit of point coincidence in the spatial coordinates $\x'\rightarrow \x$ was already taken. 
However, the dispersions described by Eq.~(\ref{i1}) are more easily calculated if we use the following integral representation of these correlation functions,
\begin{align}
\langle E_x(\x,t)E_x(\x,t')\rangle_{{}_{\tt vacuum}} &= -\frac{3}{8\pi^2}\int_{1}^\infty (u^2+1)\left[\frac{1}{(u\Delta t+2z)^4}+\frac{1}{(u\Delta t-2z)^4}\right]du\,,
\label{c1-2}\\ 
\langle E_z(\x,t)E_z(\x,t')\rangle_{{}_{\tt vacuum}} &= \frac{3}{4\pi^2}\int_{1}^\infty (u^2-1)\left[\frac{1}{(u\Delta t+2z)^4}+\frac{1}{(u\Delta t-2z)^4}\right]du\,,
\label{c2-2}
\end{align}
where the prescription $\Delta t\rightarrow \Delta t-i\epsilon$, with $\epsilon>0$, was used. Introducing Eqs.~(\ref{c1-2}) and (\ref{c2-2}) in Eq.~(\ref{i1}), and performing the integrals (see the appendix A for further details), we obtain that
\begin{align}
\frac{m^2}{q^2}\langle(\Delta v_{{}_\perp})^2\rangle &= -\frac{\tau}{32\pi^2z^3}\left\{\ln\left[\frac{(\tau-2z)^2+4\tau_s^2}{(\tau+2z)^2+4\tau_s^2}\right]+\frac{4\tau_s}{\tau}\arg\left[1+\frac{\tau^2}{4(\tau_s-iz)^2}\right]\right\},
\label{f1}
\\
\frac{m^2}{q^2}\langle(\Delta v_{{}_\parallel})^2\rangle &=\frac{1}{2}\frac{m^2}{q^2}\langle(\Delta v_{{}_\perp})^2\rangle-\frac{\tau^2\left(\tau^2-4z^2+12\tau_s^2\right)}{8\pi^2(\tau_s^2+z^2)[(\tau-2z)^2+4\tau_s^2][(\tau+2z)^2+4\tau_s^2]}\;.
\label{f2}
\end{align}
Here $\arg(w)$ is the argument of the complex number $w$. 

\subsection{Late time behaviour}
\label{latetime}
Regarding the behaviour of the dispersions at zero temperature, the above expressions generalize previous results and allows us to investigate new effects. First, due to the controllable switching duration, which brings more reality to the model, we are able to understand the origin of the residual effect appearing in previous works \cite{ford2004,delorenci2016}. The late time regime can be obtained directly from Eqs. (\ref{f1}) and (\ref{f2}) by taking the limit of $\tau\rightarrow\infty$, and yields
\begin{eqnarray}
\lim_{\tau\to\infty} \langle\left(\Delta v_{{}_\perp}\right)^2\rangle
&=& \frac{q^2}{4\pi^2 m^2 z^2}\left\{1-\frac{\tau_s}{2z}\arg\bigg[\Big(1+i \frac{z}{\tau_s}\Big)^2\bigg]\right\},
\label{vaccumperplt}
\\
\lim_{\tau\to\infty}\langle(\Delta v_{{}_\parallel})^2\rangle&=&\frac{1}{2}\lim_{\tau\to\infty} \langle\left(\Delta v_{{}_\perp}\right)^2\rangle - \frac{q^2}{8\pi^2 m^2 \tau_s^2}\frac{1}{1+(z/\tau_s)^2}\,.
\label{vacuumparalt}
\end{eqnarray}
If we additionally take the limit of $\tau_s\rightarrow 0$, we get that $\langle\left(\Delta v_{{}_\perp}\right)^2\rangle = q^2/(4\pi^2 m^2 z^2)$ and $\langle ( \Delta v_{{}_\parallel} )^2\rangle=0$  \cite{ford2004}. On the other hand, both dispersions vanish if the limit of $\tau_s\rightarrow\infty$ is taken, as it occurs when a model exhibiting an adiabatic transition is assumed \cite{delorenci2016}. 

\begin{figure}[h!]
\center
\includegraphics[scale=0.42]{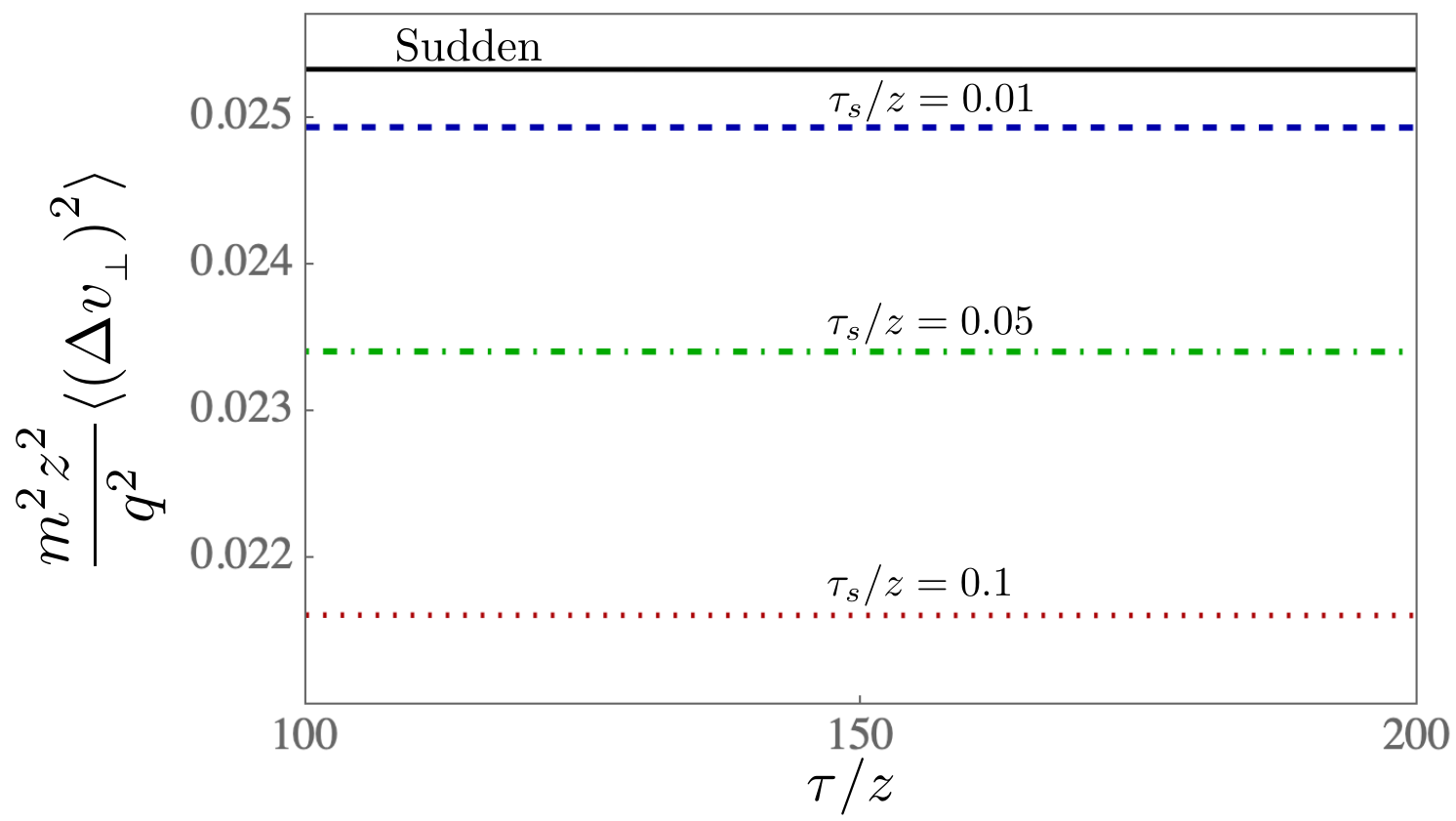}
\caption{Late time behaviour of the dispersion of the particle velocity in the perpendicular direction, described by Eq.~(\ref{vaccumperplt}), as function of $\tau/z$ for some suggestive values of the ratio $\tau_s/z$.}
\label{fig15}
\end{figure}

One important conclusion here is that the residual effect is dependent on the value of $\tau_s$, i.e., it is related to the duration of the transition between the two vacuum states. This aspect can be inspected in Fig.~\ref{fig15}, where the dispersion in the perpendicular direction is shown for some values of $\tau_s$. 
The largest possible value of this effect occurs when $\tau_s\rightarrow 0$, which  corresponds to an idealized model presenting a sudden transition. 

\subsection{Near boundary behaviour}
\label{near}
%

\begin{figure}[h!]
\center
\includegraphics[scale=0.38]{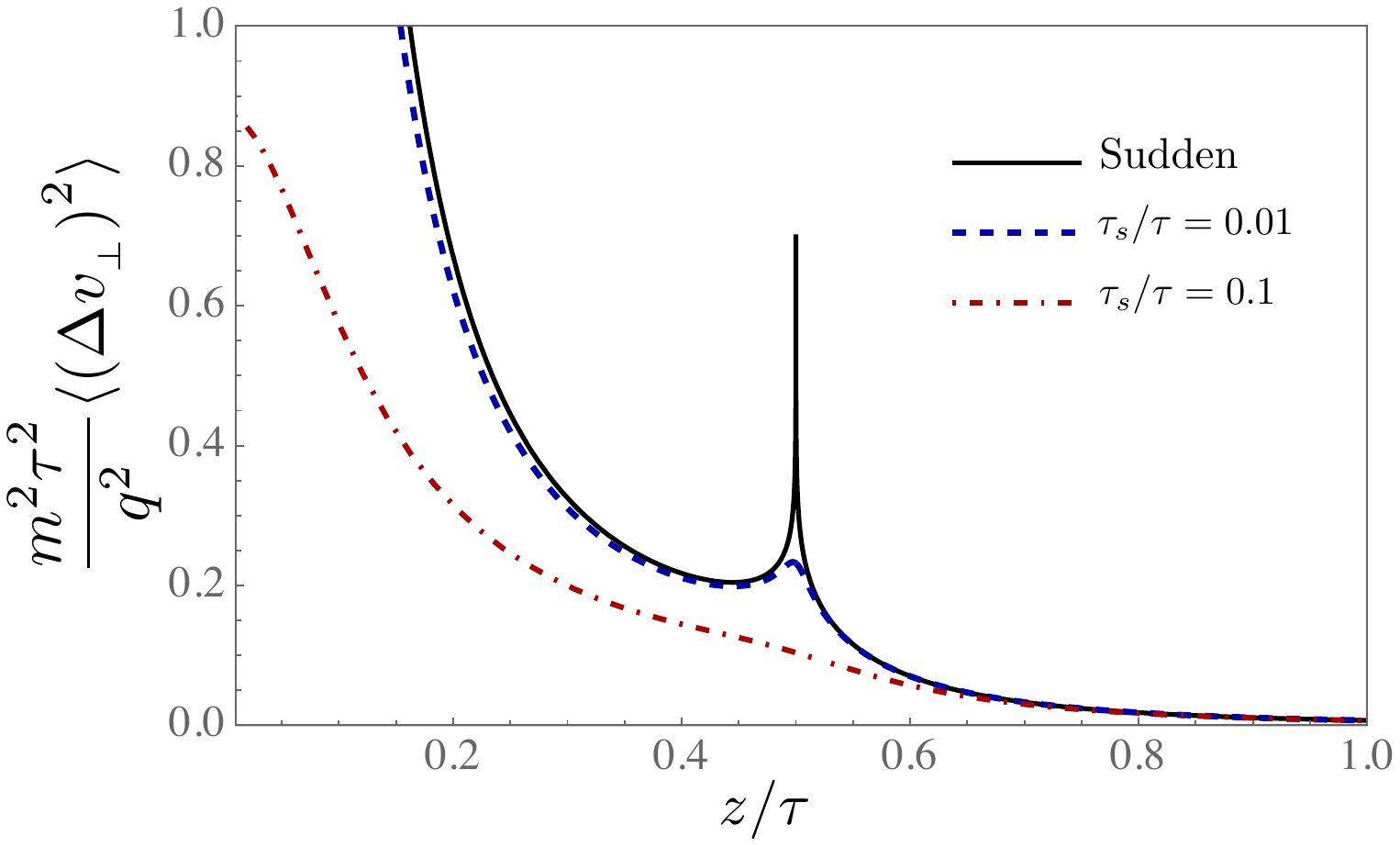}
\caption{Dispersion of the perpendicular component of the particle velocity, described by Eq.~(\ref{f1}), as function of $z/\tau$ for some suggestive values of the ratio $\tau_s/\tau$. }
\label{fig2}
\end{figure}

If we assume the regime of $z/\tau_s \ll 1$ it can be shown that the dominant contributions in Eqs.~(\ref{f1}) and (\ref{f2}) are given by
\begin{eqnarray}
\langle(\Delta v_{{}_\perp})^2\rangle
\approx \frac{q^2\tau^2}{12\pi^2 m^2 \tau_s^2}\frac{\tau^2+12\tau_s^2}{(\tau^2+4\tau_s^2)^2} \approx -\langle(\Delta v_{{}_\parallel})^2\rangle\, .
\label{ress1}
\end{eqnarray}
A finite switching transition leads to finite dispersions even at $z=0$, with magnitudes that depend on $\tau_s$. It is interesting to notice that the dispersion is negative in the parallel direction, which means that subvacuum effects may take place near the wall. This is an aspect that could not be examined in previous models because of the lack of a natural scale to compare distances. Here, this scale is provided by the duration of the transition between the vacuum states. 
%
Just to have an illustration of the effect, the distance behaviour of the perpendicular component of the dispersions is depicted in Fig.~\ref{fig2}, where some representative values of the ratio $\tau_s/\tau$ were chosen. The solid curve, which are singular at $z=0$ and $z=\tau/2$, correspond to the limit of $\tau_s\rightarrow 0$. Notice that, for finite $\tau_s$ the dispersion presents a finite magnitude on the wall.

The limit of $\tau_s\rightarrow 0$ should be taken directly in Eqs.~(\ref{f1}) and (\ref{f2}), and leads to the results predicted by the model presenting a sudden transition \cite{ford2004}. 
%
In this particular case the dispersion of the parallel component is $-{1}/({3 \pi ^2 \tau ^2})$, and the dispersion of the perpendicular component is singular on the wall.
Such singularity is a consequence of renormalizing a result that was obtained by implementing perfectly reflecting boundary conditions \cite{birrel1982}. Hence, in such an idealized model care must be taken when near to the wall behaviour is investigated. On the other hand, when the smooth transition is implemented, the dispersions are regular functions of time at any distance $z$, including $z=0$. 

Other mechanisms could be thought to regularize such divergence. We mention, for instance, the method of assuming a fluctuating boundary \cite{delorenci2014,bartolo2015}.

\subsection{Contribution to the kinetic energy due to the vacuum transition}
\label{energy}
%
Let us now study the contribution to the energy of the particle due to its interaction with the vacuum fluctuations of the quantum electric field. We stress that the effects due to the classical interaction between the particle and the conducting wall are being suppressed from our analysis. 

\begin{figure}[h!]
\center
\includegraphics[scale=0.35]{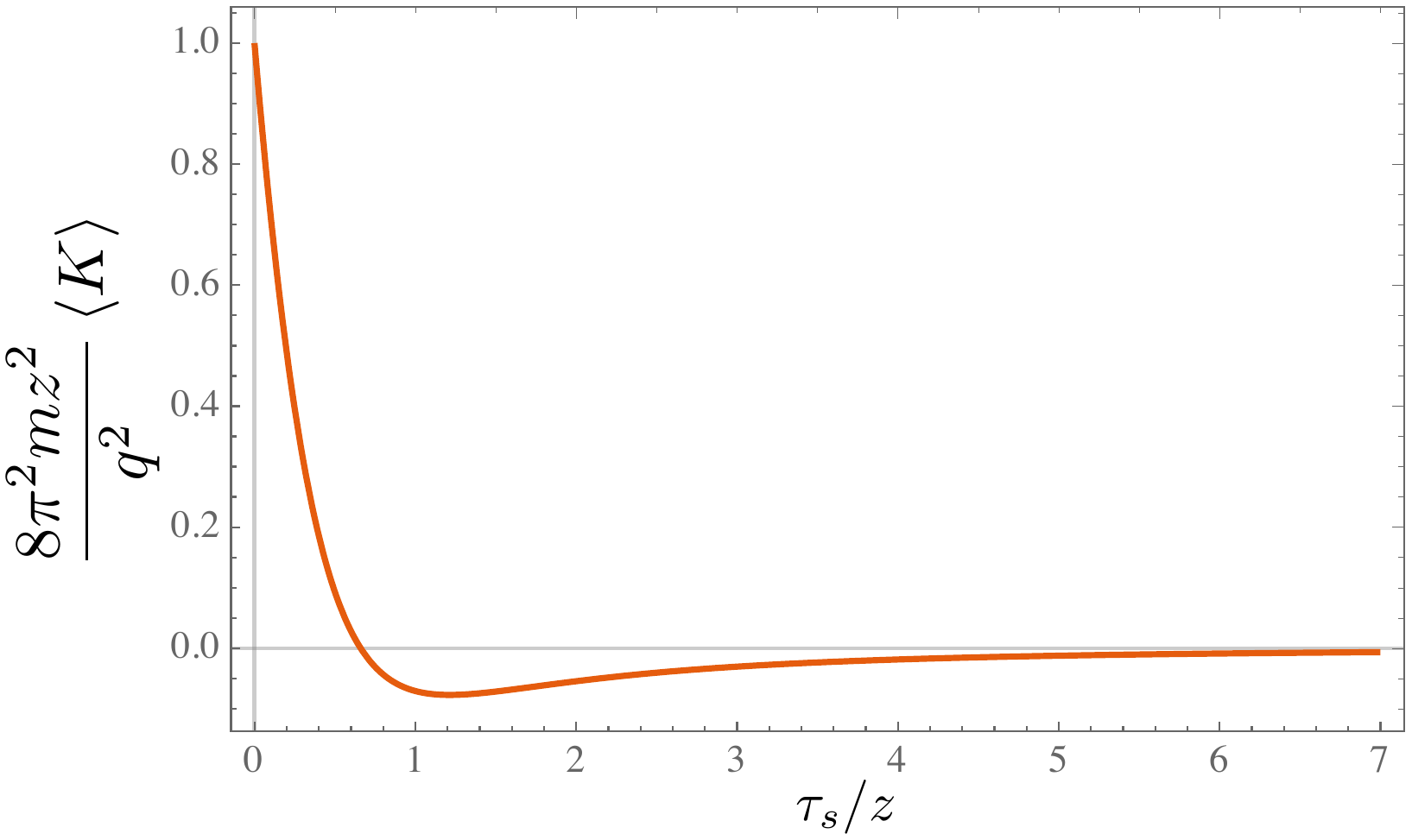}
\caption{Late time behaviour of the quantum contribution to the kinetic energy of the charged particle due to vacuum fluctuations near a conducting wall.}
\label{fig1}
\end{figure}

The dispersions derived in the last section show that there will be a contribution to the kinetic energy of the particle due to the quantum vacuum fluctuations, and it is given by $\braket{K}= m \left( \braket{{v_x}^2}+ \braket{{v_z}^2}/2\right) = m [\langle(\Delta v_{{}_\parallel})^2\rangle +  \langle\left(\Delta v_{{}_\perp}\right)^2\rangle/2]$. 
In the late time regime its main contribution can be written as,
\begin{equation}
\braket{K} = \frac{q^2}{8\pi^2 m z^2}\left[\frac{1+2(\tau_s/z)^2}{1+(\tau_s/z)^2}-2\frac{\tau_s}{z}\arctan\frac{z}{\tau_s}\right],
\label{kinetic}
\end{equation}
%
whose behaviour as function of $\tau_s/z$ is depicted in Fig.~\ref{fig1}. 
There are some aspects that should be remarked. First, the limit of $\tau_s\to 0$ leads to $\braket{K}=q^2/(8 \pi ^2 m z^2)$, which is the result obtained when the sudden switching is implemented. This corresponds to the maximum magnitude of transferred energy to the particle due to the transition between the vacuum states. 
On the other hand, when $\tau_s\to \infty$ no energy is exchanged, as it is expected in an adiabatic process. 
The kinetic energy is positive for small values of $\tau_s/z$ and gets negative values as $\tau_s/z$ gets larger values. 
There is one particular value of $\tau_s/z$ ($\approx 0.66$) for which $\braket{K}$ vanishes. What happens at this point is that negative vacuum fluctuations exactly cancel the positive vacuum fluctuations effects over the motion of the test particle. 
This means that if an experiment is prepared in such way that the switching time takes exactly this value, there will be no change in the kinetic energy of the particle sourced by the quantum vacuum fluctuations. This is an interesting result, as a finite value of $\tau_s/z$ corresponds to a non-adiabatic transition process. 
For values larger than that, subvacuum effects dominate and the kinetic energy of the particle will be lessen by a small amount whose magnitude decreases as $\tau_s/z$ increases. 

\section{Finite temperature regime}
\label{finite}
\subsection{Dispersions at finite temperature}
\label{finite-dispersions}
In this section we extend our analysis to the case where the system is in thermal equilibrium at a finite temperature $T=1/\beta$. This corresponds to exchange $\langle E_j(\x,t)E_j(\x,t')\rangle_{\tt vacuum}$ in Eq.~(\ref{i1}) by $\langle E_j(\x,t)E_j(\x,t')\rangle_{\beta}$, which is the $j$-th component of the renormalized thermal averaged correlation function \cite{brown1969}. Let us decompose this finite temperature correlation function in terms of its vacuum, thermal, and mixed parts as 
$\langle E_j(\x,t)E_j(\x,t')\rangle_{\beta} = \langle E_j(\x,t)E_j(\x,t')\rangle_{\tt vacuum}+\langle E_j(\x,t)E_j(\x,t')\rangle_{\tt thermal}+\langle E_j(\x,t)E_j(\x,t')\rangle_{\tt mixed}$. 
The vacuum term $\langle E_j(\x,t)E_j(\x,t')\rangle_{\tt vacuum}$ was already presented in Sec.~\ref{dispersions}, while the pure thermal contribution is isotropic, and reads
\begin{equation}
\langle E_j(\x,t)E_j(\x,t')\rangle_{\tt thermal} = \frac{2}{\pi^2}\mbox{Re}\sum_{n=1}^{\infty}\frac{1}{(\Delta t+in\beta)^4}\; ,
\label{c-thermal}
\end{equation}
for any $j$. Notice that, besides this contribution be the same in any direction, it does not depend on distance $z$, i.e., it is not sensitive to the presence of the wall. 
Finally, the mixed contribution presents two distinct components,
\begin{align}
\langle E_x(\x,t)E_x(\x,t')\rangle_{\tt mixed} &=-\frac{2}{\pi^2}\mbox{Re}\sum_{n=1}^{\infty}\frac{(\Delta t+in\beta)^2 +4z^2}{[(\Delta t+in\beta)^2-4z^2]^3}\;,
\label{ct2a}
\\
\langle E_z(\x,t)E_z(\x,t')\rangle_{\tt mixed} &=\frac{2}{\pi^2}\mbox{Re}\sum_{n=1}^{\infty}\frac{1}{[(\Delta t+in\beta)^2-4z^2]^2}\;.
\label{ct2b}
\end{align}
%
%
Just for completeness, we should mention that the correlations in directions parallel to the wall coincide $\langle E_y(\x,t)E_y(\x,t')\rangle_{\beta} = \langle E_x(\x,t)E_x(\x,t')\rangle_{\beta}$. 

Using the above results, the finite temperature dispersions $\langle(\Delta v_j)^2\rangle_{{}_\beta}$ can be obtained in the same way as we did for the zero temperature case. First, let us split the different contributions for each component of the dispersions as 
\begin{equation}
\langle(\Delta v_j)^2\rangle_{{}_\beta} = \langle(\Delta v_j)^2\rangle_{\tt vacuum}+\langle(\Delta v_j)^2\rangle_{\tt thermal}+\langle(\Delta v_j)^2\rangle_{\tt mixed},
\label{dbeta}
\end{equation}
and thus they can be calculated by using the above results for the correlation functions.

The pure vacuum contributions  $\langle(\Delta v_{{}_\perp})^2\rangle_{\tt vacuum}$ and $\langle(\Delta v_{{}_\parallel})^2\rangle_{\tt vacuum}$ to the total dispersions are those already presented in Eqs.~\eqref{f1} and \eqref{f2}, respectively (we have introduced the label {\tt vacuum} to these zero temperature dispersions just to make clear its meaning when compared to the finite temperature contributions).

Next, using the general formula given in Eq.~\eqref{i1}, we obtain the following result for the pure thermal contribution, 
\begin{equation}
\frac{m^2}{q^2}\langle(\Delta v_j)^2\rangle_{\tt thermal}=\frac{1}{3\pi^2\beta^2}\left[2\psi^{{}^{(1)}}\!\!\left(1+\frac{2\tau_s}{\beta}\right)-\psi^{{}^{(1)}}\!\!\left(1+\frac{2\tau_s+i\tau}{\beta}\right)-\psi^{{}^{(1)}}\!\!\left(1+\frac{2\tau_s-i\tau}{\beta}\right)\right],
\label{thermalpure}
\end{equation}
which is spatially homogeneous and isotropic. Here $\psi^{{}^{(n)}}(x)$ represent the polygamma functions \cite{gradshteyn2007},
\begin{equation}
 \psi^{{}^{(n)}}(x)=(-1)^{(n+1)}n!\sum_{k=0}^\infty \frac{1}{(k+x)^{n+1}}.
\end{equation}

Finally, the remaining dispersions $\langle(\Delta v_{{}_\perp})^2\rangle_{\tt mixed}$ and $\langle(\Delta v_{{}_\parallel})^2\rangle_{\tt mixed}$ can be handled by combining Eqs.~\eqref{c1-2}, \eqref{c2-2}, and \eqref{app2}, and using
\begin{align}
{\cal F}^{\pm}(a,b)\doteq&\int_0^1dx(1\pm x^2)\psi^{{}^{(1)}}(a+bx)
=\frac{1\pm 1}{b}\psi^{{}^{(0)}}(a+b)-\frac{1}{b}\psi^{{}^{(0)}}(a)
\nonumber \\
&\mp\frac{2}{b^2}\psi^{{}^{(-1)}}(a+b)\pm\frac{2}{b^3}\left[\psi^{{}^{(-2)}}(a+b)-\psi^{{}^{(-2)}}(a)\right],
\end{align} 
where the polygamma function $\psi^{{}^{(n)}}(x)$ for negative $n$ is studied in Ref. \cite{adamchik1998}. Hence,
\begin{align}
\frac{m^2}{q^2}\langle(\Delta v_{{}_\perp})^2\rangle_{\tt mixed}=\frac{1}{2\pi^2\beta^2}\mbox{Re}&\left[2{\cal F}^{-}\left(1+\frac{2\tau_s}{\beta},-\frac{2iz}{\beta}\right)-{\cal F}^{-}\left(1+\frac{2\tau_s+i\tau}{\beta},-\frac{2iz}{\beta}\right)
\right.
\nonumber
\\
&\left.
-{\cal F}^{-}\left(1+\frac{2\tau_s-i\tau}{\beta},-\frac{2iz}{\beta}\right)\right],
\label{mixedperp}
\\
\frac{m^2}{q^2}\langle(\Delta v_{{}_\parallel})^2\rangle_{\tt mixed}=-\frac{1}{4\pi^2\beta^2}\mbox{Re}&\left[2{\cal F}^{+}\left(1+\frac{2\tau_s}{\beta},-\frac{2iz}{\beta}\right)-{\cal F}^{+}\left(1+\frac{2\tau_s+i\tau}{\beta},-\frac{2iz}{\beta}\right)
\right.
\nonumber
\\
&\left.
-{\cal F}^{+}\left(1+\frac{2\tau_s-i\tau}{\beta},-\frac{2iz}{\beta}\right)\right].
\label{mixedparallel}
\end{align}

\subsection{Time behaviour of the dispersions}
\label{time}
\begin{figure}[h!]
\center
\includegraphics[scale=0.39]{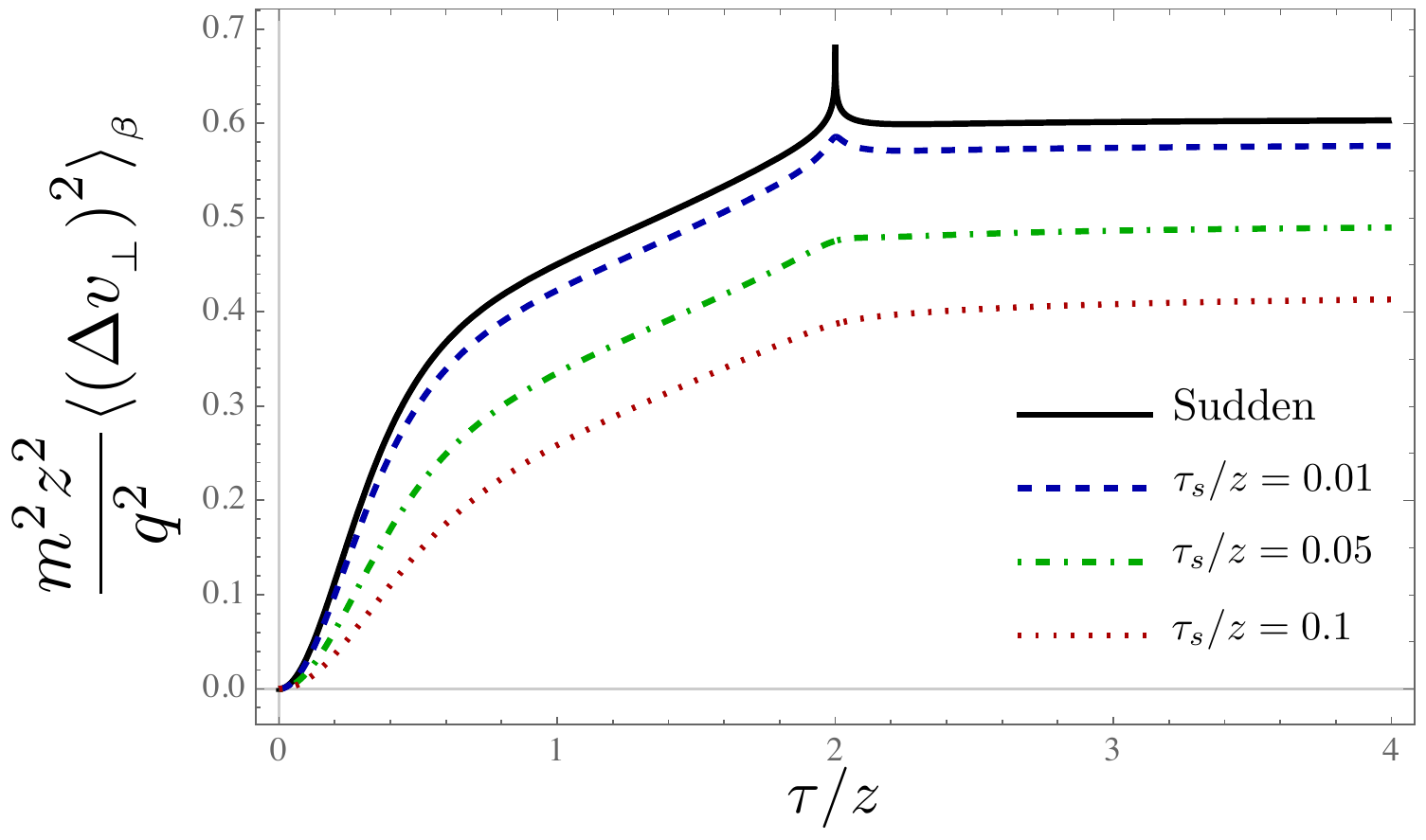}
\caption{Behaviour of the perpendicular component of the finite temperature dispersion as function of time $\tau/z$ for some values of switching time $\tau_s/z$.  In this figure we choose $\beta/z = 0.5$.}
\label{fig8}
\end{figure}

The behaviour of  $\langle(\Delta v_{{}_\perp})^2\rangle_{{}_\beta}$ is depicted in Fig.~\ref{fig8} for some chosen values of $\tau_s/z$. 
The solid curve is singular at $\tau=2z$ and corresponds to the limiting case of $\tau_s\rightarrow 0$ \cite{hongwei2006}. 
%
The curves related to the smooth transitions ($\tau_s \neq 0$) are always regular and positive, and they are always bellow the one related to the sudden transition. 
As both thermal and mixed contributions are positive, the total effect linked to the finite temperature contributions is to increase the magnitude of the dispersion, with a stronger influence in the interval $\tau<2z$.  In all cases there will be residual effects in the late time regime. 

\begin{figure}[h!]
\center
\includegraphics[scale=0.39]{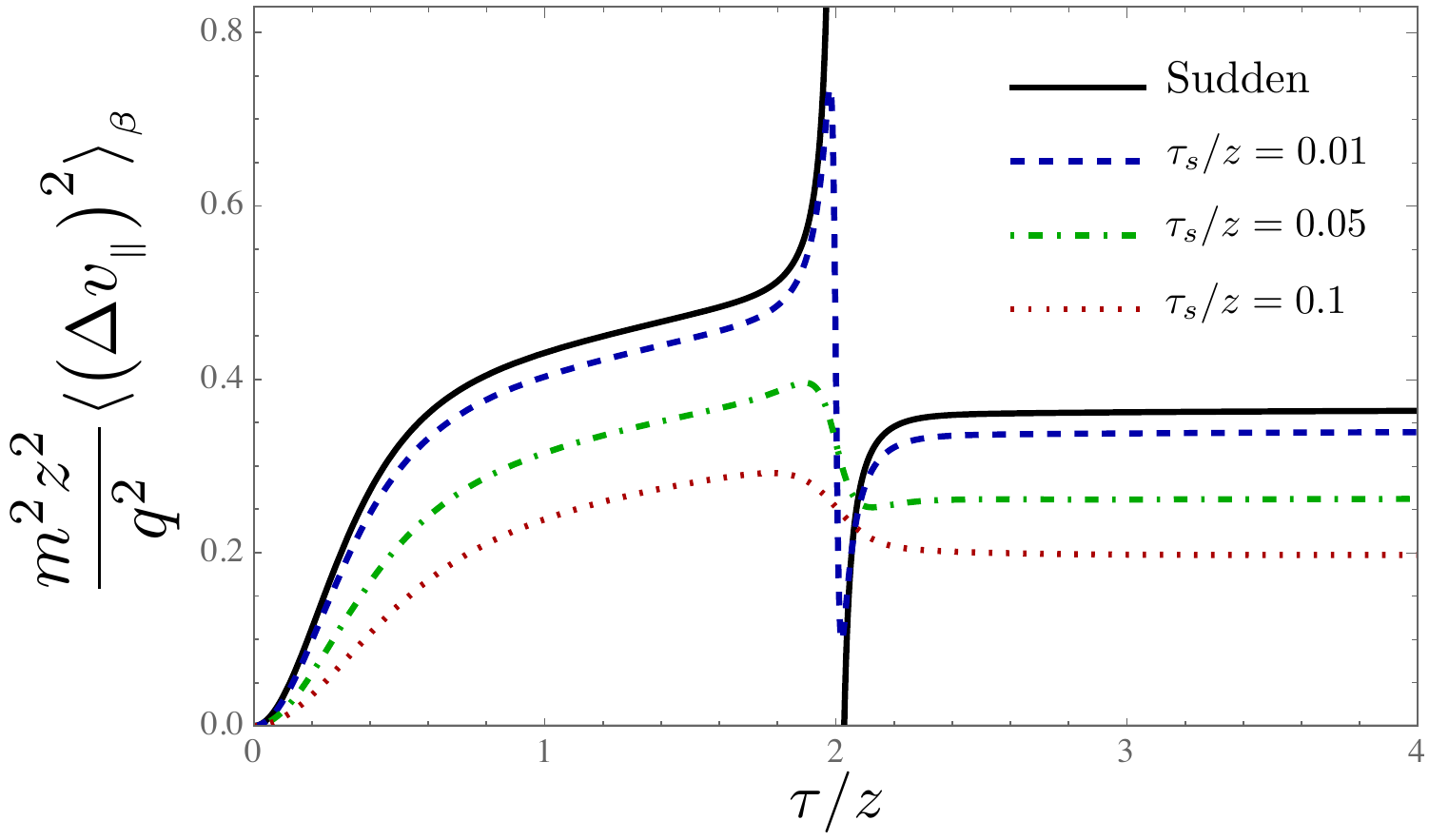}
\caption{Behaviour of the parallel component of the finite temperature dispersion as function of time $\tau/z$ for some values of switching time $\tau_s/z$.  In this figure we choose $\beta/z = 0.5$.}
\label{fig9}
\end{figure}

The behaviour of $\langle(\Delta v_{{}_\parallel})^2\rangle_{{}_\beta}$ is depicted in Fig.~\ref{fig9} for some representative values of $\tau_s/z$. 
In this case the dispersion is positive for $\tau < 2z$ and can be positive or negative for $\tau>2z$, depending on the temperature. Recall that in the regime of zero temperature this dispersion is always negative for $\tau>2z$. Now, thermal and mixed contributions present opposite signals, but for high temperature the thermal contribution is stronger and leads the dispersion to be positive. Similarly to the behaviour exhibited in the perpendicular direction, if compared to the zero temperature case, there is a distortion of the curves in the interval $\tau <2z$ caused by the thermal contribution. The solid curve describes the model presenting a sudden transition \cite{hongwei2006}.  Moreover, there will be residual effects in the late time regime. 
\begin{figure}[h!]
\center
\includegraphics[scale=0.39]{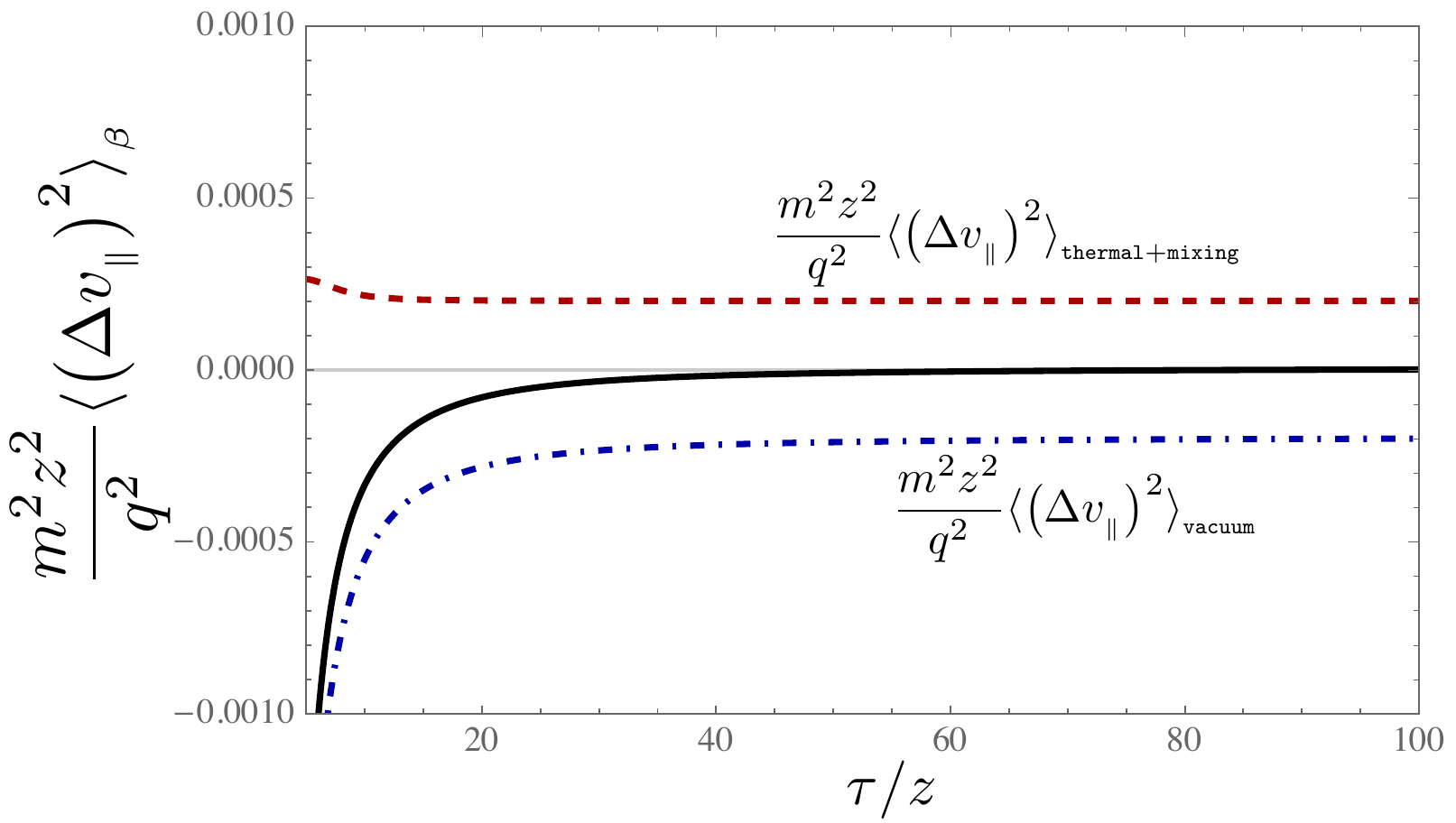}
\caption{The competition between positive thermal (dashed curve) and negative vacuum (dot-dashed curve) fluctuations can produce a vanishing component of the dispersion parallel to the wall in the late time regime. Here we set $\tau_s/z=0.01$, for which a cooling effect occurs at $\beta/z \approx 6.3$.}
\label{fig14}
\end{figure}  
However, we notice that there will be configurations such that, even at finite temperature, no residual effect will survive. This is because the positive contribution added by thermal effects can be exactly cancelled by the negative contribution coming from the vacuum term. This is a sort of quantum cooling effect produced by subvacuum fluctuations. This aspect is described in Fig.~\ref{fig14} for a specific case. When such regime is achieved only the dispersion of the velocity perpendicular to the wall survives. 

\subsection{Distance behaviour of the dispersions}
\label{distance}
\begin{figure}[h!]
\center
\includegraphics[scale=0.39]{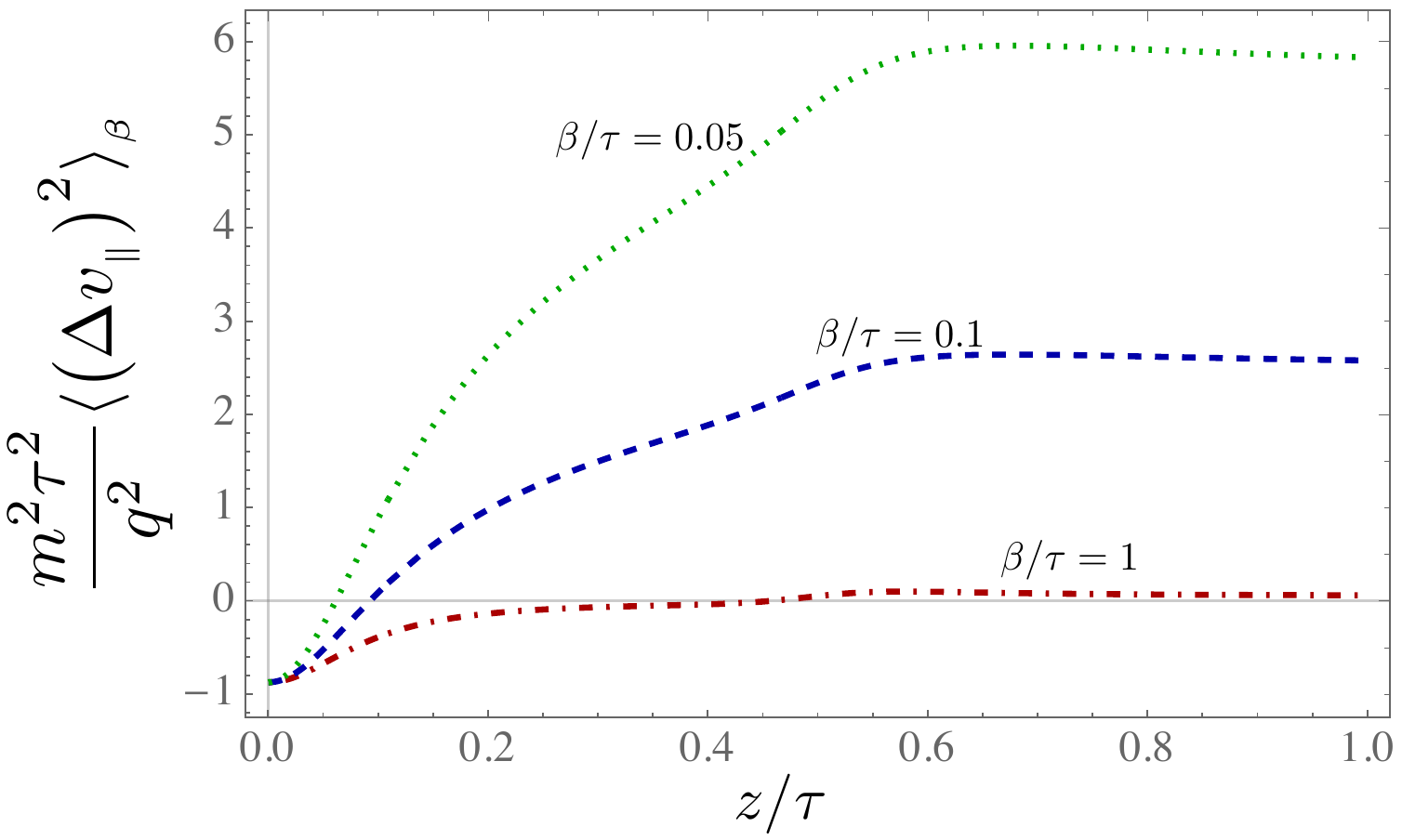}
\caption{Behaviour of the parallel component of the dispersion near the wall. Notice that the magnitude of the dispersion in the limit of $z\rightarrow 0$ does not depend on the temperature. In this figure we choose $\tau_s/\tau = 0.1$.}
\label{fig6}
\end{figure}

If we assume the regime of $z/\beta \ll1$, the main contributions to the dispersions will be 
$
\langle(\Delta v_{{}_\parallel})^2\rangle_{\tt mixed}\approx-\langle(\Delta v_j)^2\rangle_{\tt thermal}
$
and
$
\langle(\Delta v_{{}_\perp})^2\rangle_{\tt mixed}\approx\langle(\Delta v_j)^2\rangle_{\tt thermal},
$
implying that near the wall the dispersion of the parallel component of the particle velocity is sourced only by the vacuum term. This behaviour is depicted in Fig.~\ref{fig6}, where a switching time $\tau_s/\tau = 0.1$ was set. 
Notice that when $z$ goes to zero the magnitude of the dispersion converge to a value that does not depend on the temperature, as the sum of mixed and thermal contributions go to zero in this limit. The value of the dispersion at $z=0$ depends only on $\tau_s / \tau$, as discussed in Sec.  \ref{near}.
%
On the other hand, additionally to the vacuum term, thermal and mixed terms do contribute to the dispersion in the perpendicular direction. The magnitude of such influence depends on the temperature, being larger for higher temperatures. 
\begin{figure}[h!]
\center
\includegraphics[scale=0.39]{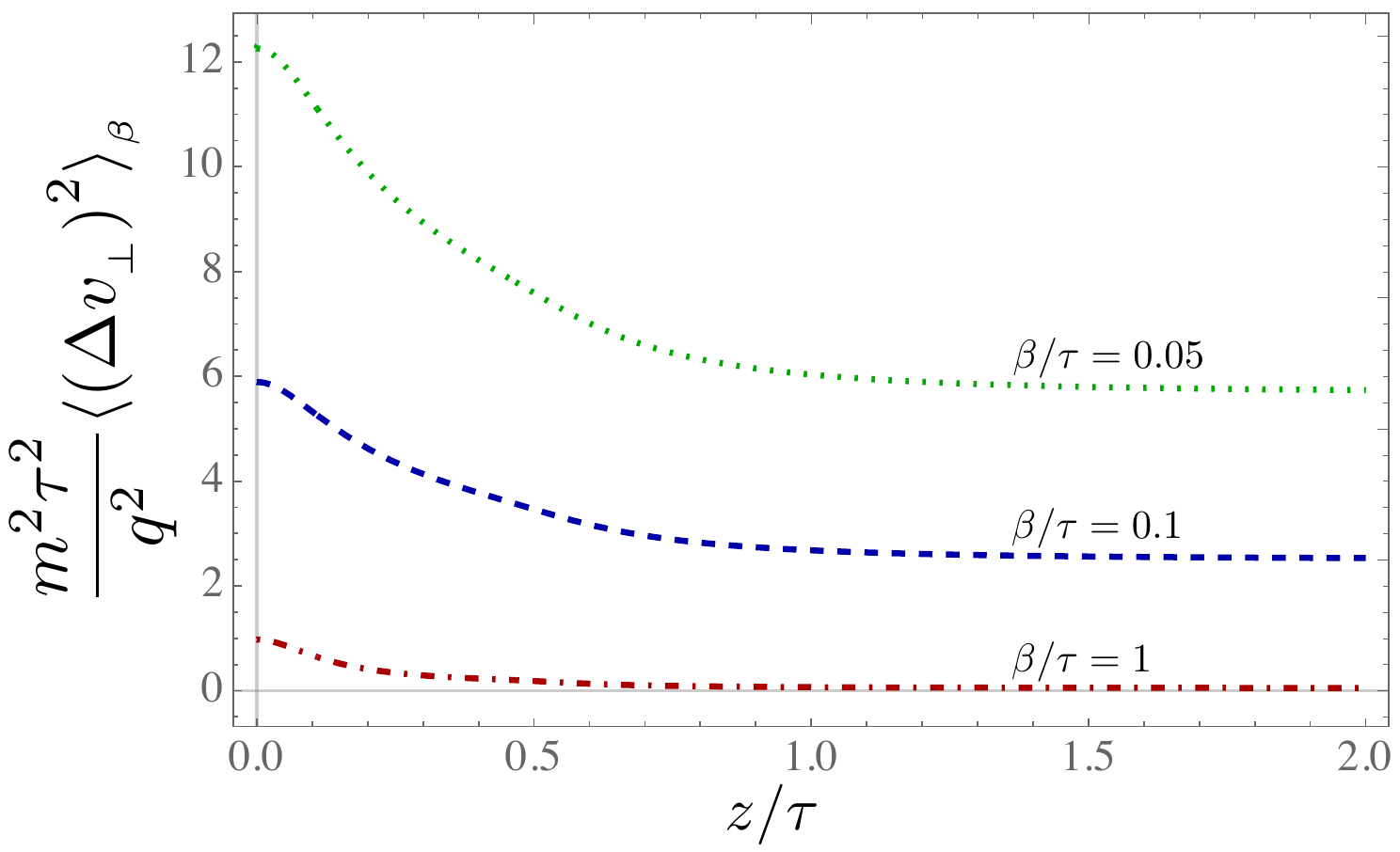}
\caption{Behaviour of the perpendicular component of the dispersion near the wall. Notice that, differently from the parallel component, now  the magnitude of the dispersion when $z\rightarrow 0$ depends on the temperature, for a same transition duration.  In this figure we choose $\tau_s/\tau = 0.1$.}
\label{fig7}
\end{figure}  
This behaviour is depicted in Fig.~\ref{fig7}, where it is clearly seen that temperature can significantly affect the dispersion arbitrarily near the wall. Moreover, it is worth noticing that the magnitude of the temperature dependent contribution to the dispersion in the perpendicular direction can even be larger than the vacuum contribution, as it can be confirmed by inspecting Figs.~\ref{fig2} and \ref{fig7} (for $\tau_s/\tau=0.1$). 
%
%


\subsection{Thermal versus vacuum dominance near the wall}
\label{dominance}
The magnitude of the effect produced by quantum fluctuations on the motion of the particle can be measured by means of the mean squared velocity, 
$$
\langle v^2\rangle_{\beta} \doteq \sum_j \langle v_j^2\rangle_{\beta} = \sum_j \langle \left(\Delta v_j\right)^2\rangle_{\beta}= 2\langle(\Delta v_{{}_\parallel})^2\rangle_{\tt \beta} + \langle(\Delta v_{{}_\perp})^2\rangle_{\tt \beta}\, ,
$$ 
where each component in the above expression can be obtained using Eq. (\ref{dbeta}).
When the limit of zero temperature is taken, we obtain the pure vacuum mean squared velocity 
$$
 \lim_{\beta\rightarrow\infty }\langle v^2\rangle_{\beta} = 2\langle(\Delta v_{{}_\parallel})^2\rangle_{\tt vacuum} + \langle(\Delta v_{{}_\perp})^2\rangle_{\tt vacuum}  \doteq  \langle v^2\rangle.
$$ 
In order to compare thermal to vacuum contributions to the dispersions we define the fractional difference parameter $\gamma$ as
\begin{equation}
\gamma = \left|\frac{\langle v^2\rangle_{\beta}-\langle v^2\rangle}{\langle v^2\rangle} \right| ,
\end{equation}
so that when $\gamma > 1$ thermal effects dominate over the vacuum effects. On the other hand when $\gamma < 1$ vacuum effects will dominate. 

We will limit our analysis to the late time regime. For this proposal, making use of the results in Eqs. (\ref{vaccumperplt}) and (\ref{vacuumparalt}), and those listed in appendix B, we obtain that
\begin{eqnarray}
\lim_{\tau\rightarrow\infty} \gamma &=&  2\,\Bigg|
\bigg[
 \left(\frac{2z}{\beta}\right)^2\psi^{{}^{(1)}}\!\!\left(1+\frac{2\tau_s}{\beta}\right) 
 + \frac{2z}{\beta}\operatorname{Im} \psi^{{}^{(0)}}\!\!\left(1+\frac{2\tau_s}{\beta}-\frac{2iz}{\beta}\right)
\nonumber\\
&& - 2\operatorname{Re}\psi^{{}^{(-1)}}\!\! \left(1+\frac{2\tau_s}{\beta}-\frac{2iz}{\beta}\right) 
- \frac{\beta}{z}\operatorname{Im} \psi^{{}^{(-2)}}\!\!\left(1+\frac{2\tau_s}{\beta}-\frac{2iz}{\beta}\right) 
\bigg]
\nonumber \\
&&\times \left\{\frac{2+(z/\tau_s)^2}{1+(z/\tau_s)^2}-\frac{\tau_s}{z}\arg\bigg[\Big(1+i \frac{z}{\tau_s}\Big)^2\bigg]\right\}^{-1}\Bigg|.
\label{gammaLT}
\end{eqnarray}

Notice that for a smooth transition ($\tau_s/\beta \neq 0$), $\gamma$ is a non null regular function of $\tau_s/\beta$ on the wall. This aspect can be seen if we evaluate the above equation on the wall, i.e.,
\begin{equation}
\lim_{z\rightarrow 0}\lim_{\tau\rightarrow\infty} \gamma = \left(\frac{4\tau_s}{\beta}\right)^2\psi^{{}^{(1)}}\!\!\left(1+\frac{2\tau_s}{\beta}\right),
\label{gammaLTz0}
\end{equation}
which is a function that vanishes when $\beta\rightarrow\infty$, as expected, but can achieve values smaller or larger than 1, depending on the ratio $\tau_s/\beta$. Specifically, if we solve Eq. (\ref{gammaLTz0}) for $\gamma = 1$ we obtain $\tau_s/\beta \approx 0.26$. This value can be used as a reference value to study the dispersions near the wall in the late time regime. Thus, using this result in Eq.~(\ref{gammaLT}) and solving it to $\gamma =1$, we obtain that $z/\beta \approx 0.015$. This means that, when $z/\beta > 0.015$, finite temperature effects will dominate over vacuum effects. In order to have an estimate, suppose the particle is placed at a distance $z=1\mu {\rm m}$ from the wall. In this case, a temperature of the order of $34 {\rm K}$ is enough to produce a thermal dominance over vacuum effects. 
This result should be compared with the one predicted by the model presenting a sudden transition \cite{hongwei2006}. In such idealized case thermal dominance at this distance from the wall would require temperatures higher than $700 {\rm K}$. 

The origin of such discrepancy relies on the use of results that are not valid arbitrarily near the conducting wall. This is the case when the idealized sudden process is implemented. In such model, the combination of perfect boundary conditions with the renormalization process introduces an artificial divergence at $z=0$, and the results should be taken with care in the regime of small distances, mainly when the system does not provide a natural scale of distance to allow comparisons. On the other hand, the regularized results provided by the model presenting a smooth transition allows us to study the effects arbitrarily near the wall, as  discussed along this work, and depicted in Figs. \ref{fig6} and \ref{fig7} for some specific cases. Here, a natural scale of distance is provided by the duration of the transition, which is always finite in a real physical system. 

\section{Final remarks}
\label{final}
The idea of implementing a smooth transition between the vacuum states of the system here studied was already reported in the literature \cite{seriu2008,seriu2009,delorenci2016}. Particularly, in Ref. \cite{delorenci2016} the transition was described by means of the normalized $C^\infty$ switching function $F^{(n)}_\tau(t) = c_n\big[1+\left(2t/\tau\right)^{2n}\big]^{-1}$, where $n$ is a positive integer that should be chosen according to the layout of the system, and $c_n = (2n/\pi)\sin(\pi/2n)$ is the normalization constant.
As a result the dispersions could be analytically integrated and their behaviour investigated as function of time and distance to the wall.
%
%
However, contrarily to the previous models, the dispersions vanish in the late time regime. This is because the switching time introduced by this function is a linear function of $\tau$.
Hence, making $\tau\rightarrow\infty$ implies an infinity transition duration, which means that there would be no change of the vacuum state.
%
%

As a further generalization of this framework, the concept of a smooth transition with a controllable switching interval of time is here introduced, and the analysis was extended to the regime of finite temperature. The main motivation is that a real system will always take a nonzero duration to evolve between the two vacuum states. The switching function describing the transition is now given by Eq.~(\ref{int27}).  All previous results regarding the behaviour of the dispersions of the particle velocity are now recovered as limiting cases of this more realistic model.
As a consequence, more than just generalizing previous models and recovering their predictions in some appropriate limits, new effects related to this important system were found. 
For instance, it was shown that thermal effects can be more important than vacuum effects arbitrarily near the wall, contrarily to the previous expectations. Additionally, the residual effect in the late time regime was here reported to be linked to the duration of the transition. In this sense, we understand that this effect is a sort of particle energy exchanging due to the vacuum states transition. Moreover, even in finite temperature regime, it was shown that the kinetic energy of the particle can be diminished due to subvacuum quantum fluctuations. 

A map connecting our findings with the results reported in the literature can be set as follows. The sudden transition at zero temperature \cite{ford2004} is obtained by taking the limits of $\tau_s\rightarrow 0$ and $\beta \rightarrow \infty$ in the dispersions given by $\langle(\Delta v_j)^2\rangle_{\beta}$. The sudden transition at finite temperature \cite{hongwei2006} is obtained by taking the limit of $\tau_s\rightarrow 0$. 
The smooth transition at zero temperature \cite{delorenci2016} is obtained by taking the limit of $\beta \rightarrow \infty$ and adjusting the switching time $\tau_s$ in terms of the parameter $n$. Particularly, in this case the late time regime leads to vanishing dispersion when we set $\tau_s\rightarrow\infty$, that describes the case where no transition occurs. 
Finally, the smooth transition at zero temperature with an independently controllable switching duration (described here in Sec.~\ref{dispersions})  is obtained by taking the limit of $\beta \rightarrow \infty$. In this case residual effects occur, whose magnitude depends on $\tau_s$.

Mathematically, functions $F^{(n)}_\tau(t)$ and $F_{\tau_s,\tau}(t)$ play a similar role in the description of a smooth transition between the vacuum states. However, while the former leads to easier integrations in the calculation of the dispersions, the later has the advantage of allowing a controllable duration of the switching. This is an important aspect when we are interested in studying the late time regime, the near to the wall effects, and the energy gained/lost by the particle due to the transition, as discussed in Sec.~\ref{energy}. The magnitude and also the signal of the contribution from the vacuum fluctuations to the total energy of the particle depends a lot on the duration of the transition $\tau_s$. As shown in Fig.~\ref{fig1}, subvacuum effects may occur for system's configurations for which $\tau_s$ is greater than a certain value, and in such cases the particle will have part of its energy suppressed by a certain amount. This is a result that can not be found in the realm of classical physics. When thermal contributions are considered, the main effect is to move the curve in Fig.~\ref{fig1} up to the positive region, leading to an increase of the total energy of the particle, as expected.

Closing, we would like to remark that we here investigated the contribution to the motion of a spinless charged particle due to vacuum fluctuations of the electromagnetic field in the presence of a reflecting wall. When classical effects are included, as for instance the interaction between the electric charge and the conducting wall, the total kinetic energy of the particle will be written as $K_{{}_{\tt total}} = K_{{}_{\tt classical}} + \langle K \rangle$, where the last term in this sum is calculated with the dispersions  examined in the previous sections. Hence, negative $\langle K \rangle$ does not mean that the kinetic energy of the particle is negative, but only that subvacuum fluctuations are able to decrease the magnitude of its energy. 




\appendix
\section{Further details in the derivation of the vacuum dispersions}
\label{appendix}
An important step in the calculations leading to Eqs.~(\ref{f1}) and (\ref{f2}) is to notice that, 
$$
\frac{d F_{\tau_s,\tau}}{dt} = \frac{1}{\pi\tau_s}\left[\frac{1}{1+t^2/\tau_s^2}-\frac{1}{1+(\tau - t)^2/\tau_s^2}\right].
$$
Thus, after introducing (\ref{c1-2}), (\ref{c2-2}) in (\ref{i1}), integrating by parts, and using residue theorem, we obtain
\begin{equation}
\int_{-\infty}^\infty\int_{-\infty}^\infty\frac{F_{\tau_s,\tau}(t)F_{\tau_s,\tau}(t')}{(\Delta t+a)^4}dtdt' = \frac{1}{6}\left[\frac{1}{(\tau-2i\tau_s+a)^2}+\frac{1}{(\tau+2i\tau_s-a)^2}-\frac{2}{(2i\tau_s-a)^2}\right].
\label{app2}
\end{equation} 
Finally, the remaining integral in $u$ variable can be solved in terms of elementary functions by using the result
\begin{equation}
\int_{1}^\infty du\frac{u^2\pm 1}{u^2(u+a)^2}=\frac{1}{1+a}\pm\left[\frac{a+2}{a^2(1+a)}-\frac{2}{a^3}\ln(1+a)\right].
\nonumber
\end{equation}

\section{Dispersions in the late time regime}
\label{appendixii}
%
%
Taking the limit of $\tau\rightarrow\infty$ in Eqs.~(\ref{thermalpure}), (\ref{mixedperp}) and (\ref{mixedparallel}), we obtain the late time expressions for the thermal,
\begin{equation}
\lim_{\tau\to\infty}\langle(\Delta v_\perp)^2\rangle_{\tt thermal}=\lim_{\tau\to\infty}\langle(\Delta v_\parallel)^2\rangle_{\tt thermal}=\frac{2q^2}{3\pi^2m^2\beta^2}\psi^{{}^{(1)}}\!\!\left(1+\frac{2\tau_s}{\beta}\right),
\nonumber
\end{equation}
and mixed contributions,
\begin{eqnarray}
\lim_{\tau\to\infty} \langle\left(\Delta v_{{}_\perp}\right)^2\rangle_{\tt mixed}
&=& -\frac{q^2}{4\pi^2 m^2 z^2}\left[
\frac{\beta}{z}\operatorname{Im} \psi^{{}^{(-2)}}\!\!\left(1+\frac{2\tau_s}{\beta}-\frac{2iz}{\beta}\right) 
+ 2\operatorname{Re}\ln \Gamma \left(1+\frac{2\tau_s}{\beta}-\frac{2iz}{\beta}\right)\right] ,
\nonumber 
\\
\lim_{\tau\to\infty}\langle(\Delta v_{{}_\parallel})^2\rangle_{\tt mixed} 
&=& \frac{1}{2}\lim_{\tau\to\infty} \langle\left(\Delta v_{{}_\perp}\right)^2\rangle_{\tt mixed}+ \frac{q^2}{2\pi^2 m^2 \beta z} \operatorname{Im} \psi^{{}^{(0)}}\!\!\left(1+\frac{2\tau_s}{\beta}-\frac{2iz}{\beta}\right).
\nonumber 
\end{eqnarray}

\acknowledgments
This work was partially supported by CNPq (Conselho Nacional de Desenvolvimento Cient\'{\i}fico e Tecnol\'ogico) under Grant No. 302248/2015-3, and FAPESP (Funda\c{c}\~ao de Amparo \`a Pesquisa do Estado de S\~ao Paulo) under grant 2015/26438-8.


\begin{thebibliography}{99}
%
\bibitem{ford2004} 
H. Yu and L. H. Ford, 
Vacuum fluctuations and Brownian motion of a charged test particle near a reflecting boundary, 
\href{https://doi.org/10.1103/PhysRevD.70.065009}{{Phys. Rev. D}, \textbf{70} (2004) 065009}
[\href{https://arxiv.org/abs/quant-ph/0406122}{arXiv:quant-ph/0406122}].

\bibitem{ford2005}
L. H. Ford, 
Stochastic spacetime and Brownian motion of test particles, 
\href{https://doi.org/10.1007/s10773-005-8893-z}{Int. J. Theor. Phys. {\bf 44} (2005) 1753}
[\href{https://arxiv.org/abs/gr-qc/0501081}{arXiv:gr-qc/0501081}].

\bibitem{hongwei2004} 
H. Yu and J. Chen, 
Brownian motion of a charged test particle in vacuum between two conducting plates, 
\href{https://doi.org/10.1103/PhysRevD.70.125006}{Phys. Rev. D {\bf 70} (2004) 125006}
[\href{https://arxiv.org/abs/quant-ph/0412010}{arXiv:quant-ph/0412010}].

\bibitem{hongwei2006} 
H. Yu, J. Chen, and P. Wu, 
Brownian motion of a charged test particle near a reflecting boundary at finite temperature,
\href{https://doi.org/10.1088/1126-6708/2006/02/058}{JHEP {\bf 02} (2006) 058}
[\href{https://arxiv.org/abs/hep-th/0602195}{arXiv:hep-th/0602195}].

\bibitem{seriu2008} 
M. Seriu and C. H. Wu, 
Switching effect on the quantum Brownian motion near a reflecting boundary, 
\href{https://doi.org/10.1103/PhysRevA.77.022107}{{Phys. Rev. A} \textbf{77} (2008) 022107}
[\href{https://arxiv.org/abs/0711.2203}{arXiv:0711.2203}].

\bibitem{seriu2009} 
M. Seriu and C. H. Wu, 
Smearing effect due to the spread of a probe particle on the Brownian motion near a perfectly reflecting boundary, 
\href{https://doi.org/10.1103/PhysRevA.80.052101}{{Phys. Rev. A} \textbf{80} (2009) 052101}
[\href{https://arxiv.org/abs/0906.5142}{arXiv:0906.5142}].

\bibitem{delorenci2016}
V.  A. De Lorenci, C. C. H. Ribeiro, and M. M. Silva, 
Probing quantum vacuum fluctuations over a charged particle near a reflecting wall,
\href{https://doi.org/10.1103/PhysRevD.94.105017}{{Phys. Rev. D} {\bf 94} (2016) 105017}
[\href{https://arxiv.org/abs/1606.09134}{arXiv:1606.09134}].

\bibitem{camargo2018}
G. H. Camargo, V.  A. De Lorenci, F. F. Rodrigues, C. C. H. Ribeiro, and M. M. Silva, 
Vacuum fluctuations of a scalar field near a reflecting boundary and their effects on the motion of a test particle, 
\href{https://doi.org/10.1007/JHEP07(2018)173}{JHEP {\bf 07} (2018) 173}
[\href{https://arxiv.org/abs/1709.10392}{arXiv:1709.10392}].

\bibitem{bessa2014}
C. H. G. Bessa, V. A. De Lorenci, and L. H. Ford, 
Analog model for light propagation in semiclassical gravity, 
\href{https://doi.org/10.1103/PhysRevD.90.024036}{Phys. Rev. D {\bf 90} (2014) 024036}
[\href{https://arxiv.org/abs/1402.6285}{arXiv:1402.6285}].

\bibitem{delorenci2019}
V. A. De Lorenci, and L. H. Ford,
Subvacuum effects on light propagation,
\href{https://doi.org/10.1103/PhysRevA.99.023852}{Phys. Rev. A {\bf 99} (2019) 023852}
[\href{https://arxiv.org/abs/1804.10132}{arXiv:1804.10132}].

\bibitem{sopova2002} 
V. Sopova and L. H. Ford, 
Energy density in the Casimir effect, 
\href{https://doi.org/10.1103/PhysRevD.66.045026}{Phys. Rev. {\bf D66} (2002) 045026}
[\href{https://arxiv.org/abs/quant-ph/0204125}{arXiv:quant-ph/0204125}].

\bibitem{sopova2005} 
V. Sopova and L. H. Ford, 
Electromagnetic field stress tensor between dielectric half-spaces, 
\href{https://doi.org/10.1103/PhysRevD.72.033001}{Phys. Rev. {\bf D72} (2005) 033001}
[\href{https://arxiv.org/abs/quant-ph/0504143}{arXiv:quant-ph/0504143}].

\bibitem{ford1978} 
L. H. Ford, 
Quantum coherence effects and the second law of thermodynamics, 
\href{https://doi.org/10.1098/rspa.1978.0197}{Proc. R. Soc. {\bf A364} (1978) 227}.

\bibitem{ford1991} 
L. H. Ford, 
Constraints on negative-energy fluxes, 
\href{https://doi.org/10.1103/PhysRevD.43.3972}{Phys. Rev.  {\bf D43} (1991) 3972}.

\bibitem{ford1997} 
L. H. Ford and T. A. Roman, 
Restrictions on negative energy density in flat spacetime, 
\href{https://doi.org/10.1103/PhysRevD.55.2082}{Phys. Rev. {\bf D55} (1997) 2082}
[\href{https://arxiv.org/abs/gr-qc/9607003}{arXiv:gr-qc/9607003}].

\bibitem{flanagan1997} 
E. E. Flanagan, 
Quantum inequalities in two-dimensional Minkowski spacetime, 
\href{https://doi.org/10.1103/PhysRevD.56.4922}{Phys. Rev. {\bf D56} (1997) 4922}
[\href{https://arxiv.org/abs/gr-qc/9706006}{arXiv:gr-qc/9706006}].

\bibitem{fewster1998} 
C. J. Fewster and S. P. Eveson, 
Bounds on negative energy densities in flat spacetime, 
\href{https://doi.org/10.1103/PhysRevD.58.084010}{Phys. Rev. D  {\bf 58} (1998) 084010}
[\href{https://arxiv.org/abs/gr-qc/9805024}{arXiv:gr-qc/9805024}]. 

\bibitem{bessa2016}
C. H. G. Bessa, V. A. De Lorenci, L. H. Ford, and C. C. H. Ribeiro, 
Model for lightcone fluctuations due to stress tensor fluctuations,
\href{https://doi.org/10.1103/PhysRevD.93.064067}{Phys. Rev. D {\bf 93} (2016) 064067}
[\href{https://arxiv.org/abs/1602.03857}{arXiv:1602.03857}].

\bibitem{brown1969} 
L. S. Brown and G. J. Maclay, 
Vacuum Stress between conducting plates: an image solution,
\href{https://doi.org/10.1103/PhysRev.184.1272}{Phys. Rev. {\bf 184} (1969) 1272}.

\bibitem{birrel1982}
N. D. Birrel and P. C. W. Davies, {\em Quantum Fields in Curved Space}
(Cambridge University Press, Cambridge, 1982), Sec. 4.3.

\bibitem{delorenci2014}
V.  A. De Lorenci, E. S. Moreira, Jr., and M. M. Silva, 
{Quantum Brownian motion near a point-like reflecting boundary,}
\href{https://doi.org/10.1103/PhysRevD.90.027702}{{Phys. Rev. D} {\bf 90} (2014) 027702}
[\href{https://arxiv.org/abs/1404.3115}{arXiv:1404.3115}].

\bibitem{bartolo2015}
N. Bartolo, S. Butera, M. Lattuca, R. Passante, L. Rizzuto, and S. Spagnolo,
Vacuum Casimir energy densities and field divergences at boundaries,
\href{https://doi.org/10.1088/0953-8984/27/21/214015}{{J. Phys.: Condens. Matter} {\bf 27} (2015) 214015}
[\href{https://arxiv.org/abs/1410.1492}{arXiv:1410.1492}].

\bibitem{gradshteyn2007}
I. S. Gradshteyn and I. M. Ryzhik,
{\it Table of Integrals, Series, and Products}, 7th Ed. 
(Academic, New York, 2007), Eqs. 8.363-8 and 8.365-1.

\bibitem{adamchik1998}
V. S. Adamchik, 
Polygamma functions of negative order, 
\href{https://doi.org/10.1016/S0377-0427(98)00192-7}{J. Comp. Appl. Math. {\bf 100} (1998) 191}.

\end{thebibliography}
\end{document}